\documentclass[preprint,12pt]{elsarticle}
\usepackage{graphicx}
\usepackage{amssymb} 
\usepackage{amsmath}    
\usepackage{latexsym}   
\usepackage{bm}
\usepackage{times}

\def\abr{\mathrel{\stackrel{\longrightarrow}{A\!B}}\!}
\def\bar{\mathrel{\stackrel{\longrightarrow}{B\!A}}\!}

\newtheorem{prop}{Proposition}
\begin{document}
\begin{frontmatter}

\title{Distinguishing between evidence and its explanations \\ in the steering of
atomic clocks}

\author[jm]{John M. Myers\corref{cor1}}
\ead{myers@seas.harvard.edu}
\author[hm]{F. Hadi Madjid}
\ead{gmadjid@aol.com}
\cortext[cor1]{Corresponding author}
\address[jm]{Harvard School of Engineering and Applied  
Sciences, Cambridge, MA 02138, USA}
\address[hm]{82 Powers Road, Concord, MA 01742, USA}

\begin{abstract}
Quantum theory reflects within itself a separation of evidence from
explanations.  This separation leads to a known proof that: (1) no wave
function can be determined uniquely by evidence, and (2) any chosen
wave function requires a guess reaching beyond logic to things
unforeseeable.  Chosen wave functions are encoded into
computer-mediated feedback essential to atomic clocks, including
clocks that step computers through their phases of computation and
clocks in space vehicles that supply evidence of signal propagation
explained by hypotheses of spacetimes with metric tensor fields.

The propagation of logical symbols from one computer to another
requires a shared rhythm---like a bucket brigade.  Here we show how
hypothesized metric tensors, dependent on guesswork, take part in the
logical synchronization by which clocks are steered in rate and
position toward aiming points that satisfy phase constraints, thereby
linking the physics of signal propagation with the sharing of logical
symbols among computers.

Recognizing the dependence of the phasing of symbol arrivals on
guesses about signal propagation transports \emph{logical
  synchronization} from the engineering of digital communications to a
discipline essential to physics.  Within this discipline we begin to
explore questions invisible under any concept of time that fails to
acknowledge unforeseeable events.  In particular, variation of
spacetime curvature is shown to limit the bit rate of logical
communication.
\end{abstract} 

\end{frontmatter}
 
 
\maketitle

\section{Introduction}
While outcomes are subject to quantum uncertainty, uncertainty is only
the tip of an iceberg: how can one ``know'' that a wave function
describes an experimental situation?  The distinction within quantum
theory between linear operators and probabilities implies a gap
between any explanation and the evidence explained
\cite{ams02,aop05,tyler07,CUP}:
\begin{prop} \label{prop:one} To choose a wave function
to explain experimental evidence requires reaching beyond logic based
on that evidence, and evidence acquired after the choice is made can
call for a revision of the chosen wave function.
\end{prop}  
Because no wave function can be unconditionally known, not even
probabilities of future evidence can be unconditionally foreseen.
Here we show implications of the unknowability of wave functions for
the second as a unit of measurement in the International System (SI),
implications that carry over to both digital communications and to the
use of a spacetime with a metric tensor to explain clock readings
at the transmission and reception of logical symbols.

For reasons including quantum uncertainty, not even the best atomic
clocks tick quite alike; they drift in frequency and position.  Here
we develop implications of the necessity of continually adjusting
clocks in response to evidence of deviations from an aiming point,
where the aiming point depends on provisional hypotheses---i.e.,
guesswork subject to revision as prompted by accumulated evidence.
Although frequency instabilities approaching $10^{-18}$
shrink the leeway within which clock adjustments are made
\cite{hinkley}, adjustments within whatever leeway persists remain
indispensable.  Clocks that generate Universal Coordinated Time (UTC)
are steered toward aiming points that depend on both a chosen wave
function and an hypothesized metric tensor field of a curved
spacetime.  Like the chosen wave function, the hypothesis of a metric
tensor, while constrained, cannot determined by measured data.

Examining how guesses enter the operations of atomic clocks, we
noticed ubiquitous computational machinery, operating in a rhythmic
cycle.  Within this machinery, hypotheses are coded into computational
processes that interact in a feedback loop that responds to evidence,
leading to the generation of more evidence.  The machinery updates
records that determine an aiming point, and so involves the writing
and reading of records.  The writing must take place at a phase of a
cycle distinct from a phase of reading, with a separation between the
writing and the reading needed to avoid a logical short circuit.

To illustrate how physical clocks depend on computational machinery,
Sec.~\ref{sec:clk} sketches the operation of an atomic clock in which
computer-mediated feedback steers an active oscillator in frequency.
First, off line, an hypothesis about how to steer the oscillator in response to
evidence of scattering of the oscillator's radiation by one or more
passive resonant atoms is developed.  Then that hypothesis,
though developed off line on the blackboard, so to speak, is encoded
into a program in the computer memory that adjusts the oscillator on
the workbench.

 In Sec.~\ref{sec:turing} we picture an explanation used in the
 operation of a clock as a string of characters written on a tape
 divided into squares, one symbol per square.  The tape is part of a
 Turing machine modified to be stepped by a clock and to communicate
 with other such machines and with keyboards and displays.  We call
 this modified Turing machine an \emph{open machine}. The computations
 performed by an open machine are open to an inflow numbers and
 formulas incalculable prior to their entry.

Because an open machine (or indeed any digital computer) cycles through
distinct phases of memory use, the most direct propagation of symbols from
one computer to another requires a symbol from one computer to arrive
during a suitable phase of the receiving computer's cycle.  In
Sec.~\ref{sec:phasing} we elevate this phase constraint to a principle that
defines the \emph{logical synchronization} necessary to a \emph{channel}
that connects clock readings at transmission of symbols to clock readings
at their reception

Provisional hypotheses, involving guesswork about the atoms of a clock
and about signal propagation, are essential to symbol-bearing channels
between computers.  The recognition that unforeseeable evidence can
prompt revision of these hypotheses raises several types of questions
as topics for a discipline of \emph{logical synchronization} within
physics, outlined in Sec.~\ref{sec:patterns}.  The first type of
question concerns patterns of channels that are possible aiming
points, as determined in a blackboard calculation that assumes a
theory of signal propagation.  Sec.~\ref{sec:typeI} addresses examples
of constraints on patterns of channels under various hypotheses of
spacetime curvature, leading to ``phase stripes'' in spacetime
that constrain the channels possible between one open machine and
another.  An example of a freedom to guess an explanation within a
constraint of given channels is characterized by a subgroup of a group of
clock adjustments.

Sec.~\ref{sec:adj} briefly addresses the two other types of questions,
pertaining not to \emph{hypothesizing} possible aiming points `on the
blackboard', but to \emph{using} hypothesized aiming points, copied
into feedback-mediating computers, for the steering of drifting
clocks.  To model drift, we draw on quantum uncertainty relations to
express the looseness of the relation between clocks as
general-relativistic worldlines and any evidence obtainable from
physical clocks.  After discussing steering toward aiming points
copied from the blackboard, we note occasions that invite revision of
a hypothesized metric tensor and of patterns of channels chosen as
aiming points.

Sec. \ref{sec:discussion} suggests giving up `global time' 
with its predictability, in favor of attention to
logical synchronization.  A few topics attractive for
future investigation are noted.

\section{Computer-mediated feedback within a single atomic
  clock}\label{sec:clk}
\begin{quote} 
The fact is that time as we now generate it is dependent upon defined
origins, a defined resonance in the cesium atom, interrogating
electronics, induced biases, timescale algorithms, and random
perturbations from the ideal. Hence, at a significant level, time---as
man generates it by the best means available to him---is an
artifact. Corollaries to this are that every clock disagrees with
every other clock essentially always, and no clock keeps ideal or
``true'' time in an abstract sense except as we may choose to define
it \cite{allan87}.
\end{quote}

Within any atomic clock computer-mediated feedback is essential.  As
is well known, in both cesium clocks that realize the SI second and in
the most stable optical atomic clocks, the atom or atoms in the atomic
clock are passive---they do not ``tick''---so the clock needs an active
oscillator and other components in addition to the atom(s).  An atomic
clock operates a feedback loop in which a guessed hypothesis steers
the rate of ticking of its oscillator.  An atomic clock's components
include
\cite{opClk2005,opClk,F1_2002,F1_2005,F1op2005,F2}:
\begin{enumerate}
\item an active \emph{oscillator} radiating at microwave or optical
  frequencies, cycling through phases adjustable over a narrow range
  of frequency (picture a driven pendulum).
\item a controllable ``gear box,'' called a frequency \emph{synthesizer}, 
 that produces an output frequency at a variable ratio to that of the
  oscillator,
\item one or more passive resonant atoms illuminated by radiation from
the oscillator;
\item a real-time computer that controls the oscillator and the synthesizer;
\item detectors of the unforeseeable outcomes of interaction of the
  atom(s) with the oscillator's radiation that write records into the
  computer memory;
\item a formula encoded in the computer memory that defines the how
  the computer steers the oscillator frequency and the synthesizer in
  response to deviations of accumulating recorded evidence from an
  hypothesized \emph{aiming point}.
\end{enumerate}
In designing an atomic clock to realize the SI second, one encounters,
among others, the following two problems.  (a) The resonance exhibited
by the atom or atoms of the clock varies with the details of the
clock's construction and the circumstances of its operation; in
particular the resonance shifts depending on the intensity of the
radiation of the atoms by the oscillator.  (b) The oscillator,
controlled by, in effect, a knob, drifts in relation to the knob
setting.

Problem (a) is dealt with by introducing a wave function parametrized
by radiation intensity and whatever other factors one deems relevant.
The second is then ``defined'' in terms of the resonance the ``would
be found'' at zero temperature (implying zero radiation)
\cite{jentschura,DopFountain,thermal}.  For a clock using cesium 133
atoms, this imagined resonance is declared by the General Conference
of Weights and Measures to be 9 192 631 770 Hz, so that the SI second
is that number of cycles of the radiation corresponding to that
imagined resonance \cite{sp330}.

Problem (b) is dealt with by feedback that adjusts a ``knob''
that controls the oscillator, in response to detections of scattering
of the oscillator's radiation by the atom or atoms of the clock, so
that the oscillator is steered toward an aiming point at which the
detection rate is sensitive to small displacement of the oscillation from
the aiming point.

\begin{figure}[h]
\includegraphics{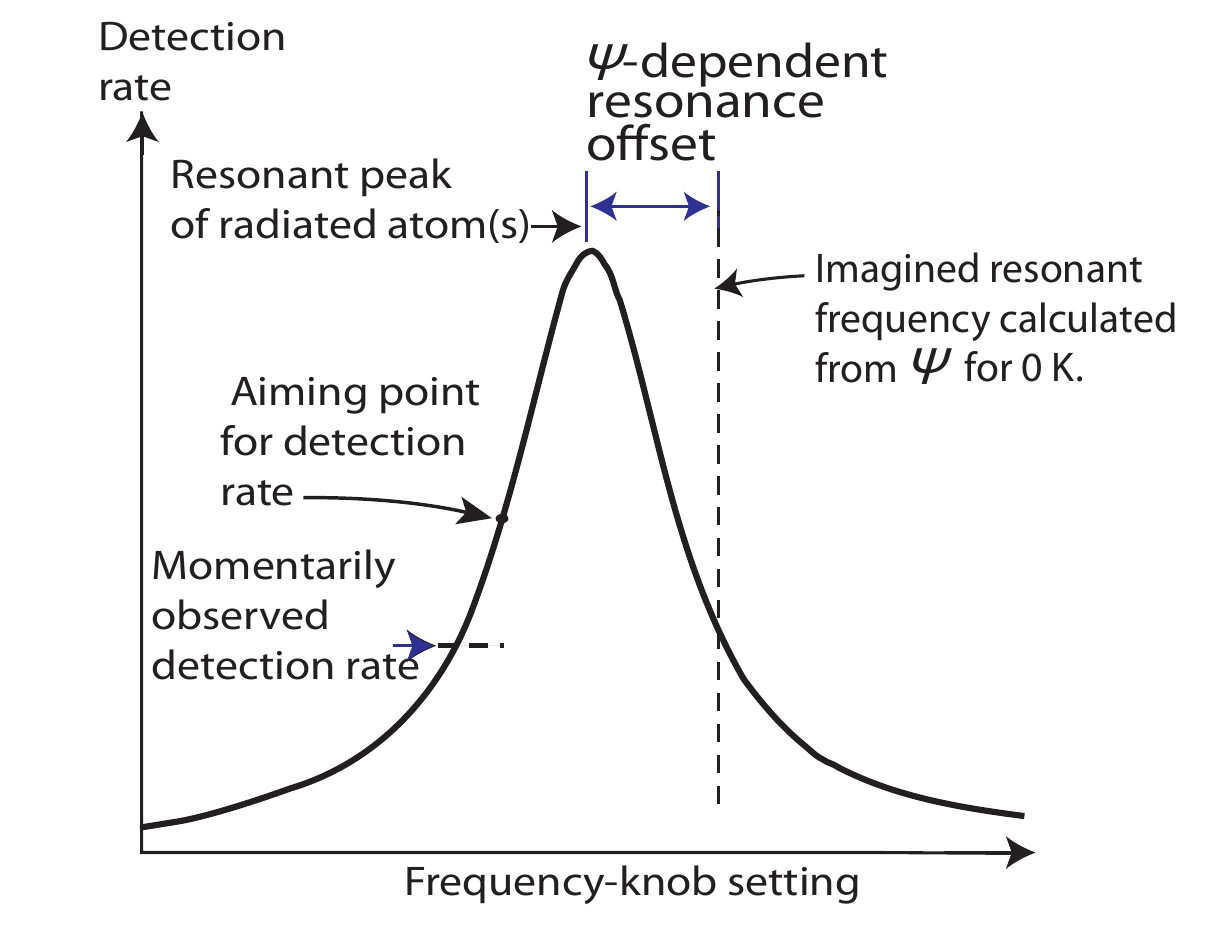}
\caption{Detection rate vs. oscillator frequency extracted from
  measurements of a clock (solid curve). Resonant frequency
  calculated for $^{133}$Cs at 0 K from chosen wave function (dashed
  line).  Feedback steers oscillator frequency $\nu_{\rm osc.}$ toward
  an aiming point for which the detection rate is sensitive
  to a small change in the oscillator frequency. \label{fig:mut}}
\end{figure} 
As illustrated in Fig.~\ref{fig:mut}, the drift of the oscillator is
indicated by variation in the detection rate (which, apart from some
statistical variation, varies with changes in the oscillator frequency
relative to the resonance of the radiated atoms).  The aiming point is
set at a fixed detection rate, chosen to be sensitive to variation in
the oscillator frequency relative to the resonance of the atoms as
radiated by the oscillator.  The function illustrated by the
bell-shaped curve, which would be obtained from experimental data, is
coded into the computer to express detection rate
\emph{vs}.\ deviation of the oscillation from the resonance of the
radiated atom(s).  The aiming point for the oscillator frequency
differs from the imagined frequency at 0 K for two reasons: (a) it
differs from the resonant peak for the atom(s) radiated by the
oscillator to get a more sensitive response, and (b) that resonant
peak differs by an amount depending on a chosen wave function from the
defining imagined resonance at 0 K.  The synthesizer gears the output
frequency of the clock to the oscillator so as to account for the
ratio of the resonance hypothesized at 0 K to the aiming point for the
oscillator.

\section{Open Turing machine as a model of a computer in a feedback loop}\label{sec:turing}
Computer-mediated feedback used in an atomic clock requires logic open
to an inflow of inputs beyond the reach of calculation.  To model the
logic of a computer that communicates with the other devices in a
feedback loop, we modify a Turing machine to communicate with external
devices, including other such machines.  One thinks of a Turing
machine, modified or not, as making a record on a tape marked into
squares, each square holding one character of an alphabet.  Operating
in a cyclic sequence of `moments' interspersed by `moves', at any
moment the machine scans one square of the tape, on which it can read
or write a single character. A move as defined in the mathematics of
Turing machines consists (only) of the logical relation between the
machine at one moment and the machine at the next moment
\cite{turing}, thus expressing the logic of a computation, detached
from its speed, so that that two computations executing at different
speeds can be represented in their logic by the same sequence of
moves.  In a feedback loop, however, computational speed matters, and
so we let the moves of the modified Turing machine be stepped by ticks
of a clock.  A step occurs once per period of revolution of the clock
hand.  This period is adjustable, on the fly. A cycle of the modified
Turing machine corresponds to a unit interval of the readings of its
clock.

To express communication between Turing machines, we postulate that
the modified Turing machine can receive externally supplied signals
and can transmit signals, with both the reception and the transmission
geared to the cycle of the machine.  In addition, the modified Turing
machine registers a count of moments at which signals are received and
moments at which signals are transmitted.  At a finer scale,
\textit{the machine records a phase quantity in the cycle of its
  clock, relative to the center of the moment at which a signal
  carrying a character arrives.}  We call such a machine an \emph{open
  machine}.  An open machine can receive detections and can command
action, for instance the action of increasing or decreasing the
frequency of the variable oscillator of an atomic clock.

In contradistinction to an open machine, one might speak of the usual
Turing machine (which Turing called an automatic machine \cite{turing})
as \emph{closed}.  Calculations performed on a closed machine proceed
from start to halt by a succession of moves made according to a
pre-programmed rule, closed to outside influences, neither receiving
anything nor commanding any action.  Such ``closed'' calculations
correspond to logic in which atomic propositions never change their
truth values.  In contrast, calculations performed on an open machine
communicating with detectors and actuators proceed by moves made
according to a rule that can be modified from outside the machine in
the course of its operation.  These calculations can respond to received
influences, such as occurrences of outcomes underivable from the
contents of the machine memory, so that the open machine writes
commands on a tape read by an external actuator.  The wider physical
world shows up in an open machine as both (1) unforeseeable messages
from external devices and (2) commands to external devices.

We picture the records made by a real-time computer in a feedback loop
as written on the tape of an open machine.  The segmentation into
moments interspersed by moves is found not just in Turing machines but
in any digital computer, which implies
\begin{prop}
The logical result of any computation is oblivious to variations in
speed at which the clock steps the computer.
 \end{prop}
\textbf{Corollary 2.1}. \textit{No computer can sense directly any variation in
its clock frequency.}
\vspace*{8pt}

Although it cannot directly sense variation in the tick rate of its clock,
the logic of open machine, thought of as stepped by an atomic clock, can
still control the adjustment of the clock's oscillator by responding to
variations in the detection rate written moment by moment onto its Turing
tape.  A flow of unforeseeable detections feeds successive computations of
results, each of which, promptly acted on, impacts probabilities of
subsequent occurrences of outcomes, even though those subsequent outcomes
remain unforeseeable.  The computation that steers the oscillator depends
not just on unforeseeable inputs, but also on a steering formula
coded into a program. \vspace*{6pt}

\noindent\textbf{Remarks}:
\begin{enumerate}
\item To appreciate feedback, one needs to distinguish any formula as
  as written from what it expresses.  For example a formula written
  along a stretch of a Turing tape as a string of characters can name
  a wave function $\psi$ that depends on a time variable $t$.  Like a
  formula chalked on a blackboard, the formula containing $\psi$, once
  written, ``sits motionless,'' in contrast to the time variation that
  the formula expresses.
\item Although unchanged over some cycles of a feedback loop, a steering
  formula does not stay put for ever.  A feedback loop operates in a larger
  context, in which steering formulas are subject to evolution.  Sooner or
  later, the string of characters that expresses the steering formula is
  apt to be overwritten by a characters expressing a new formula.
  Occasions for rewriting steering formulas are routine in clock networks,
  including those employed in geodesy and astronomy.
\item Noticing feedback raises an opportunity to improve the
  short-term stability of an atomic clock.  Recall that the oscillator
  is steered in frequency by responding to deviations in detection
  rate from an assumed operating point.  If the clock keeps a record
  of these deviations, the record could be used to make corrections to
  the clock's readings that takes the deviations into account, thereby
  improving the corrected stability of the clock.
\end{enumerate}

\section{Communication channels and logical synchronization}\label{sec:phasing} 
Because open machines (and computers) are stepped through phases by
clocks, signals communicating logical symbols from one open machine to
another must arrive at a computer  during a certain phase and
not during other phases.  A reading $\zeta_A$ of the clock of an open
machine $A$---an $A$-reading---has the form $m.\phi_m$ where an
integer $m$ indicates the count of cycles and $\phi_m$ is the phase
within the cycle.  Thus the clock reading of an open machine passes
through an integer value as the phase of the clock hand passes through
zero.  We adopt the convention that $-1/2<\phi_m\le 1/2$.  We define a
\emph{channel} from $A$ to $B$, denoted $\abr$\,, as a set of pairs,
each pair of the form $(m.\phi_m,n.\phi_n)$.  The first member
$m.\phi_m$ is an $A$-reading at which machine $A$ can transmit a
signal and $n.\phi_n$ is a $B$-reading at which the clock of machine
$B$ can register the reception of the signal.  A \emph{repeating
  channel} is defined to be a channel $\abr$ such that
\begin{equation}
  (\forall \ell \in 
[\ell_1,\ell_2])(\exists m,n,j, k) (m+\ell j.\phi_{A,\ell},n+\ell k.\phi_{B,\ell})
\in \abr,
\end{equation}
For theoretical purposes, it is convenient to define an
\emph{endlessly repeating channel} for which $\ell$ ranges over all integers.
Again for theoretical purposes, we sometimes consider channels
for which the phases are all zero, in which case one may omit writing the
phases.

When two-way repeating channels $\abr$ and $\bar$\, link
open machines $A$ and $B$, a lower bound on the clock
reading at which $A$ can receive an acknowledegment from $B$ comes
from the following
\begin{quote}
  \textbf{Definition of echo count:} Suppose that at its reading $m.0$
  an open machine $A$ transmits a signal at to an open machine $B$,
  and the first signal that $B$ can transmit back to $A$ after
  receiving $A$'s signal reaches $A$ at $m'.\phi'$; then the quantity
  $m'.\phi'- m$ will be called the echo count $\Delta_{ABA}$ at $m$.
\end{quote}
Thus $\Delta_{ABA}(m)+m$ is a lower bound on the cycle count at which
$A$ can receive an acknowledgement of a transmission sent at
$A$-reading $m$ to $B$. In the theoretical case in which receptions
occur at null phases, echo counts are integers.  Echo count is defined
relative to the variably geared output of the adjustable clock of an
open machine.

\begin{figure}[h]
\centerline{\includegraphics[width=5.2in]{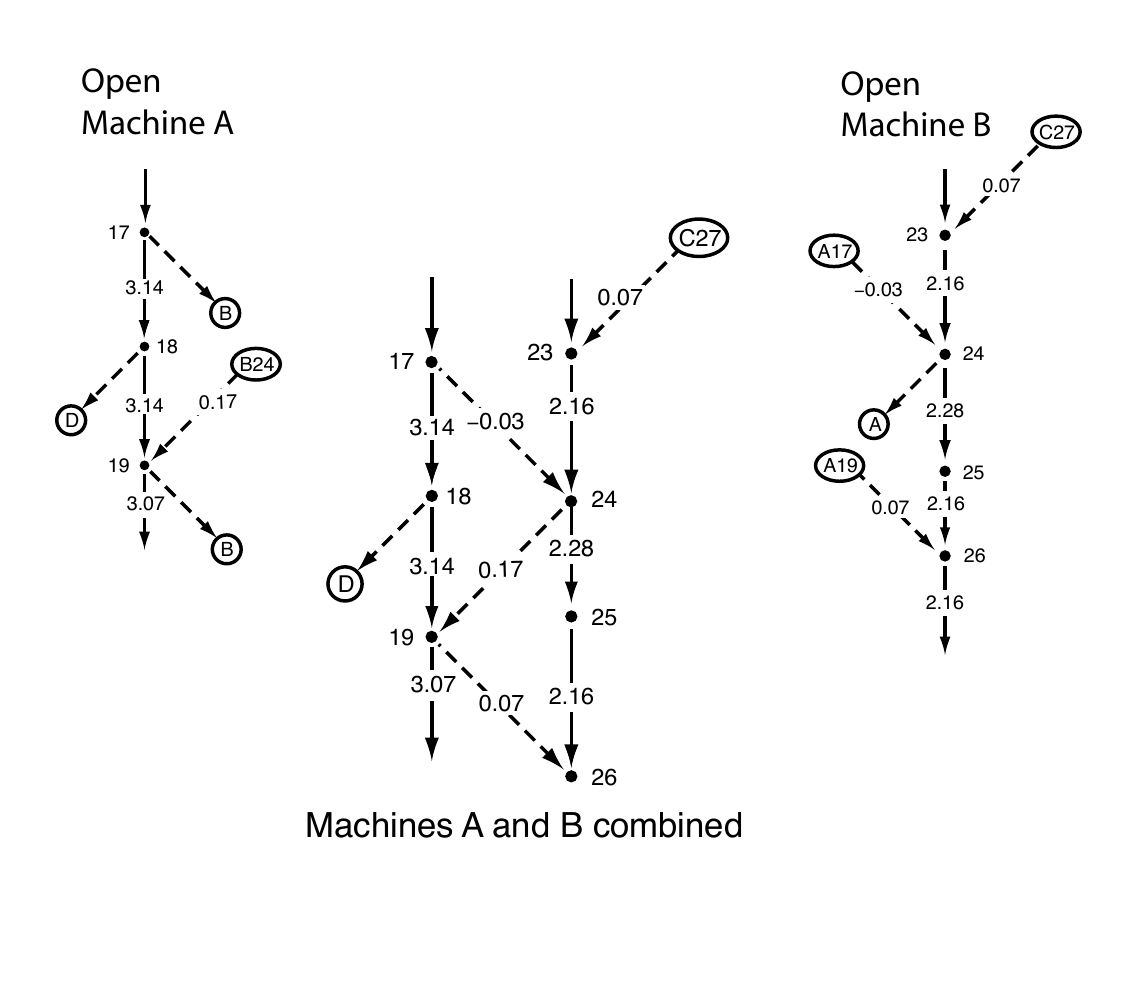}}
\caption{Graph fragments for records of open machines~A and~B
  combined; edges for open machines carry clock rate relative to the
  clock's own oscillator and incoming signal arrows (dashed) are
  labeled by phase of reception.}\label{fig:1}
\end{figure}

\begin{table}[h!]
  \begin{center}
\begin{tabular}{||ccccc||}\hline
A's cycle & Event & Other:& Phase & Cycle\\ count&&party&or
rate&sent\\ \hline \vdots &\vdots&\vdots&\vdots&\vdots\\\hline 17 &
send &B & & \\ &rate&&3.14&\\\hline 18&send&D&&\\ 
&rate&&3.14&\\\hline
19&rec'd&B&0.17&24\\ &rate&&3.07&\\ &send&B&&\\\hline
\vdots&\vdots&\vdots&\vdots&\vdots\\\hline
\end{tabular}
\end{center}
\caption{History recorded in the memory of open machine A,
indicating clock rates relative to oscillator and phases at receptions}
\label{table:one}
\end{table}
Because they are defined by local clocks without reference to any
metric tensor, channels invoke no assumption about a metric or even a
spacetime manifold.  For this reason evidence from the operation of
channels is independent of any explanatory assumptions involving a
manifold with metric and hence is independent of any global time
coordinate or any ``reference system'' \cite{soffel03}. Thus clock
readings at the transmission and the reception of signals can prompt
revisions of hypotheses about a metric tensor field.  Table
\ref{table:one} illustrates evidence in the form of clock readings
associated with channels acquired by an open machine $A$.  Histories
recorded or imagined in the form of Table \ref{table:one} can be
expressed as \emph{occurrence graphs} \cite{holtOcc}, specialized to
exhibit a distinct trail for each open machine, with the trails linked
by edges for signals.  Fig.~\ref{fig:1}, illustrates open machines
$A$ and $B$ separately, along with their combined evidence.
When ``analog'' measurements of phases with their idiosyncrasies are
forgotten, the occurrence graph for a network of endlessly repeating
channels can be ``wrapped around'' to form a marked graph
\cite{mk_graph,1639}.  Occurrence graphs, marked graphs, and more
general Petri nets \cite{peterson81} form categories with interesting
graph morphisms.

From the beating of a heart to the bucket brigade, life moves in
phased rhythms.  It is well known that for a symbol-carrying signal
transmitted from one computer to be written into the memory of a
another computer, the signal must arrive during a phase in which
writing can take place, and the cycle must offer room for a distinct
other phase.  We elevate this engineering commonplace to a principle
pertaining to open machines:
\begin{prop}\label{prop:three} A logical symbol can propagate from one
open machine to another only if the symbol arrives within the writing
phase of the receiving machine; in particular, respect for phasing
requires that for some positive $\eta$ any arrival phase $\phi_n$
satisfy the inequality 
\begin{equation}\label{eq:main} |\phi_n| < (1-\eta)/2.  
\end{equation} 
\end{prop} 
Prop. \ref{prop:three} serves as a fixed point to hold onto while
hypotheses about signal propagation in relation to channels are
subject to revision.  We call the phase constraint on a channel
asserted by (\ref{eq:main}) \emph{logical synchronization}.  

In this report we consider only channels that preserve the order of
signals in the sense that when two successive signals propagate from
any machine $A$ to any other machine $B$, whichever signal is sent
later must also arrive later:
\begin{quote}
\textbf{Order preservation:} A channel $\abr$\, preserves order if for
  any $(m.\phi_m,n.\phi_n)$ and $(m'.\phi_{m'},n'.\phi_{n'})$, 
  $m'.\phi_{m'} > m.\phi \Rightarrow n'.\phi_{n'}>n.\phi_n$.
\end{quote}
\noindent\textbf{Remarks:}
\begin{enumerate}
\item Note that $\phi_n$ in (\ref{eq:main})  is a phase of a cycle of a
  variable-rate clock \emph{not} assumed to be in any fixed
  relation to a proper clock as conceived in general relativity.
  Indeed, satisfying (\ref{eq:main}) usually requires the operation of
  clocks at variable rates.
\item Computers are commonly designed with buffering that detaches the
  timing of message reception from the stepping of the computer.
  Buffering, while convenient, inserts delay between transmission and
  the arrival of symbols in the computer memory \cite{meyr}.  In
  analyzing open machines we focus on the most direct communication
  possible, which cannot be buffered, and so employs the
  character-by-character phase meshing as asserted in
  Prop. \ref{prop:three}.
\item Logical synchronization differs from Einstein synchronization
\cite{perlick}, among other ways, by allowing leeway in the arrival of a signal.
\item Designs for logical synchronization that arise in engineering
contexts are an extensive subject \cite{meyr}.
\end{enumerate}

\section{A discipline of logical synchronization within physics}\label{sec:patterns}
Given the definition of a channel and the condition (\ref{eq:main})
essential to the communication of logical symbols, three types of
questions \vspace*{4pt} arise:

\noindent\textbf{Type I:} What patterns of interrelated channels and echo
counts can one try for as aiming \vspace*{4pt} points?

\noindent\textbf{Type II:} How can the steering of open machines be
arranged to approach given aiming points within acceptable
phase \vspace*{4pt} tolerances?

\noindent\textbf{Type III:} How to respond to deviations from aiming points
beyond
\vspace{4pt}  tolerances?\\
\noindent Such questions point the way to what might be called a
\emph{discipline of logical synchronization} transported from the
engineering of digital communications into physics. So far we
notice two promising areas of application within this discipline:
\begin{enumerate}
\item Provide a theoretical basis for networks of logically synchronized
  repeating channels, highlighting 
  \begin{enumerate}
  \item possibilities for channels with null receptive phases as a
    limiting case of desirable behavior, and
\item circumstances that force non-null phases.
  \end{enumerate}
\item Explore constraints on receptive phases imposed by
  gravitation, as a path to exploring and measuring gravitational
  curvature, including slower changes in curvature than those searched
  for by the Laser Gravitational Wave Observatory \cite{ligo}.
\end{enumerate}

\subsection{Geometry of signal propagation.}
Answers to questions of the above Types require hypotheses, if only
provisional, about signal propagation.  For this section we assume that
propagation is described by null geodesics in a Lorentzian 4-manifold $M$
with one or another metric tensor field $g$, as in general relativity.
Following Perlick \cite{perlick} we represent an open machine as a timelike
worldline, meaning a smooth embedding $\gamma\colon \zeta \mapsto
\gamma(\zeta)$ from a real interval into $M$, such that the tangent vector
$\dot{\gamma}(\zeta)$ is everywhere timelike with respect to $g$ and
future-pointing.  We limit our attention to worldlines of open machines
that allow for signal propagation between them to be expressed by null
geodesics.  To say this more carefully, we distinguish the \emph{image} of
a worldline as a submanifold of $M$ from the worldline as a mapping.
Consider an open region $V$ of $M$ containing a smaller open region $U$,
with $V$ containing the images of two open machines $A$ and $B$, with the
property that every point $a$ of the image of $A$ restricted to $U$ is
reached uniquely by one future-pointing null geodesic from the image of $B$
in $V$ and by one past-pointing null geodesic from the image of $B$ in
$V$. And suppose that this works the other way around for every point $b$
of the image of $B$ restricted to $U$.  We then say $A$ and $B$ are
\emph{radar linkable} in $U$.  We limit our attention to open machines that
are radar linkable in some spacetime region $U$.  In addition we assume
that the channels preserve order.  Indeed, we mostly deal with open
machines in a gently curved spacetime region, adequately described by Fermi
normal coordinates around a timelike geodesic.

For simplicity and to allow comparing conditions for phasing with
conditions for Einstein synchronization, we take the liberty of allowing
transmission to occur at the same phase as reception, so that both occur
during a phase interval satisfying (\ref{eq:main}).  The perhaps more
realistic alternative of demanding reception near values of $\phi=1/2$ can
be carried out with little difficulty.

To develop the physics of channels, we need to introduce three
concepts. 
\begin{enumerate}
\item We define a \emph{group of clock
  adjustments} as transformations of the readings of the clock of an
open machine.  As it pertains to endlessly repeating channels, a group
$H$ of clock adjustments consists of functions on the real numbers
having continuous, positive first derivatives.  Group multiplication
is the composition of such functions, which, being invertible, have
inverses.  To define the action of $H$ on clock readings, we speak
 `original clock readings' as distinct from 'adjusted readings'
An adjustment $f_A \in H$ acts by changing every original reading
$\zeta_A$ of a clock $A$ to an adjusted reading $f_A(\zeta_A)$.  As we
shall see, clock adjustments can affect echo counts.
\item To hypothesize a relation between the $A$-clock and an
accompanying proper clock, one has to assume one or another metric
tensor field $g$, relative to which to define proper time increments
along $A$'s worldline; then one can posit an adjustment $f_A$ such that
$f_A(\zeta_A)=\tau_A$ where $\tau_A$ is the reading imagined for the
accompanying proper clock when $A$ reads $\zeta_A$.
\item We need to speak of positional relations between open
machines. For this section we assume that when an open machine $B$
receives a signal from any other machine $A$ then $B$ echoes back a
signal to $A$ right away, so the echo count $\Delta_{ABA}$ defined
in Sec.~\ref{sec:phasing} involves no delay at $B$.  In this case,
evidence in the form of an echo count becomes explained, under the
assumption of a metric tensor field $g$, as being just twice the radar
distance \cite{perlick} from $A$ to the event of reception by $B$.
\end{enumerate}

\section{Type-I questions: mathematical expression of possible 
patterns of channels}\label{sec:typeI} Questions of Type I concern
constraints on channels imposed by the physics of signal propagation.  Here
we specialize to constraints on channels imposed by spacetime metrics,
constraints obtained from mathematical models that, while worked out so to
speak on the blackboard, can be copied onto Turing tapes as aiming points
toward which to steer the behavior of the clocks of open machines.
Questions of Types II and III are deferred to the Sec. \ref{sec:adj}.

\subsection{Channels with null phases as aiming points: two open
machines linked by a two-way channel.} We begin by considering just two
machines.  Assuming an hypothetical spacetime $(M,g)$, suppose that machine
$A$ is given as a worldline parametrized by its clock readings: what are
the possibilities and constraints for an additional machine $B$ with
two-way repeating channels $\abr$ and $\bar$ linking $B$ to $A$ at constant
echo count? We assume the idealized case of channels with null phases,
which implies integer echo counts.  For each $A$-tick there is a future
light cone and a past light cone.
\begin{figure}[h]
\centerline{\includegraphics[width=4.9in]{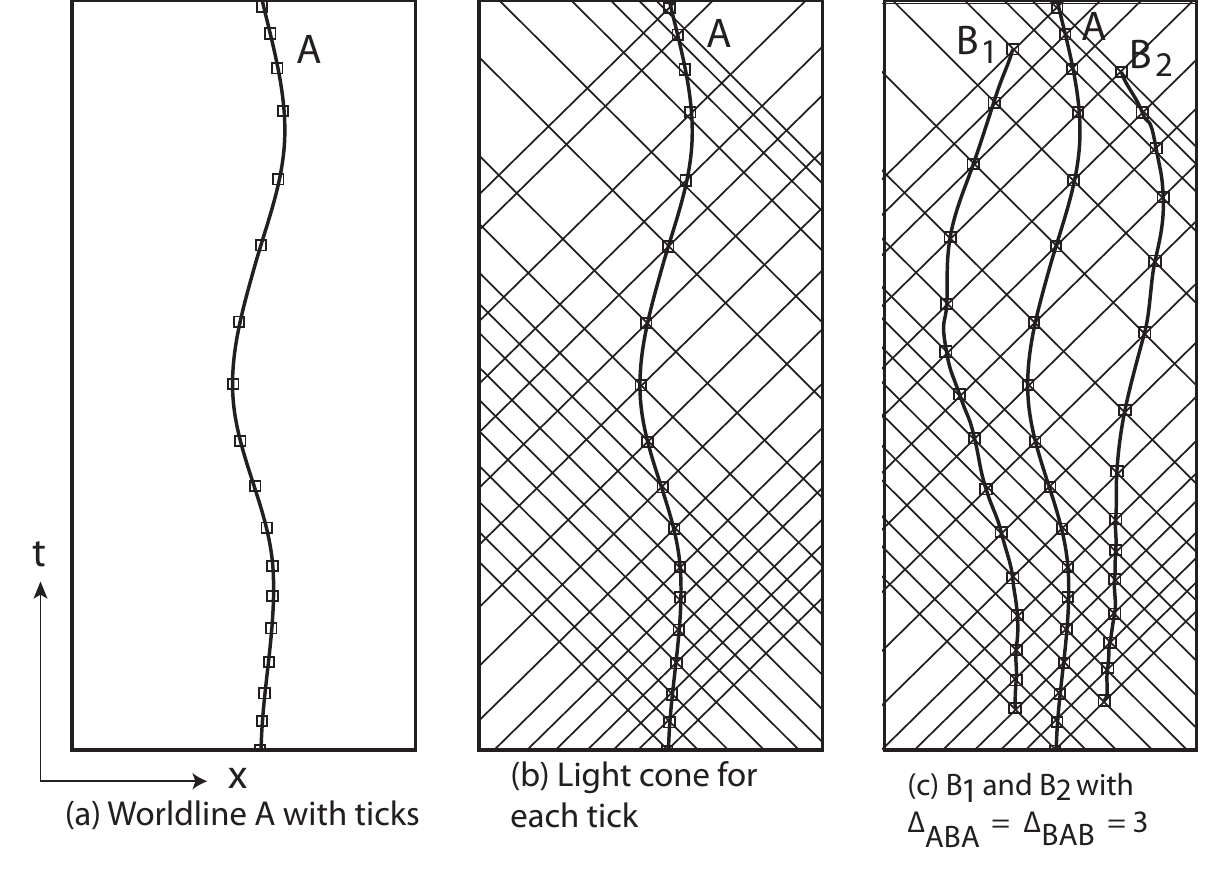}}
\caption{(a) Worldline of $A$ with tick events indicated; (b) Light
  cones associated to ticks of $A$; (c) Ticks of $B_1$ and $B_2$ at light cone
  intersections corresponding to $\Delta_{ABA}=\Delta_{BAB}=3$.
}\label{fig:3}
\end{figure}
The future light cone from an $A$-reading
$\zeta_A=m$ has an intersection with the past light cone for the
returned echo received at $\zeta_A=m+\Delta_{ABA}$.  Fig.~\ref{fig:3}
illustrates the toy case of a single space dimension in a flat
spacetime by showing the two possibilities for a machine $B$ linked to
$A$ by two-way channels at a given constant echo count.  In each
solution, the clock rate of $B$ is adjusted so that a tick of $B$ occurs at
each of a sequence of intersections of outgoing and incoming light
cones from and to ticks of $A$.  Note that the image of $B$, and not
just its clock rate, depends on the clock rate of $A$.

Determination of the tick events for $B$ leaves undetermined the $B$
trajectory between ticks, so there is a freedom of choice.  One can
exercise this freedom by requiring the image of $B$ to be consistent
with additional channels of larger echo counts. A clock adjustment of
$A$ of the form $\zeta_A\rightarrow \zeta'_A=N\zeta_A$ for $N$ a
positive integer increases the density of the two-way channel by $N$
and inserts $N-1$ events between successive $B$-ticks, thus
multiplying the echo count by $N$.  As $N$ increases without limit,
$B$ becomes fully specified.

Turning to two space dimensions, the image of $B$ must lie in a tube
around the image of $A$, as viewed in a three-dimensional space
(vertical is time).  So the image of any timelike worldline within the tube will
do for the image of $B$.  For a full spacetime of 3+1 dimensions, the
solutions for the image of $B$ fall in the corresponding
``hypertube.'' The argument does not depend on flatness and so works
for a generic, gently curved spacetime in which the channels have the
property of order preservation.

\begin{figure}[h]
\centerline{\includegraphics[width=4.9in]{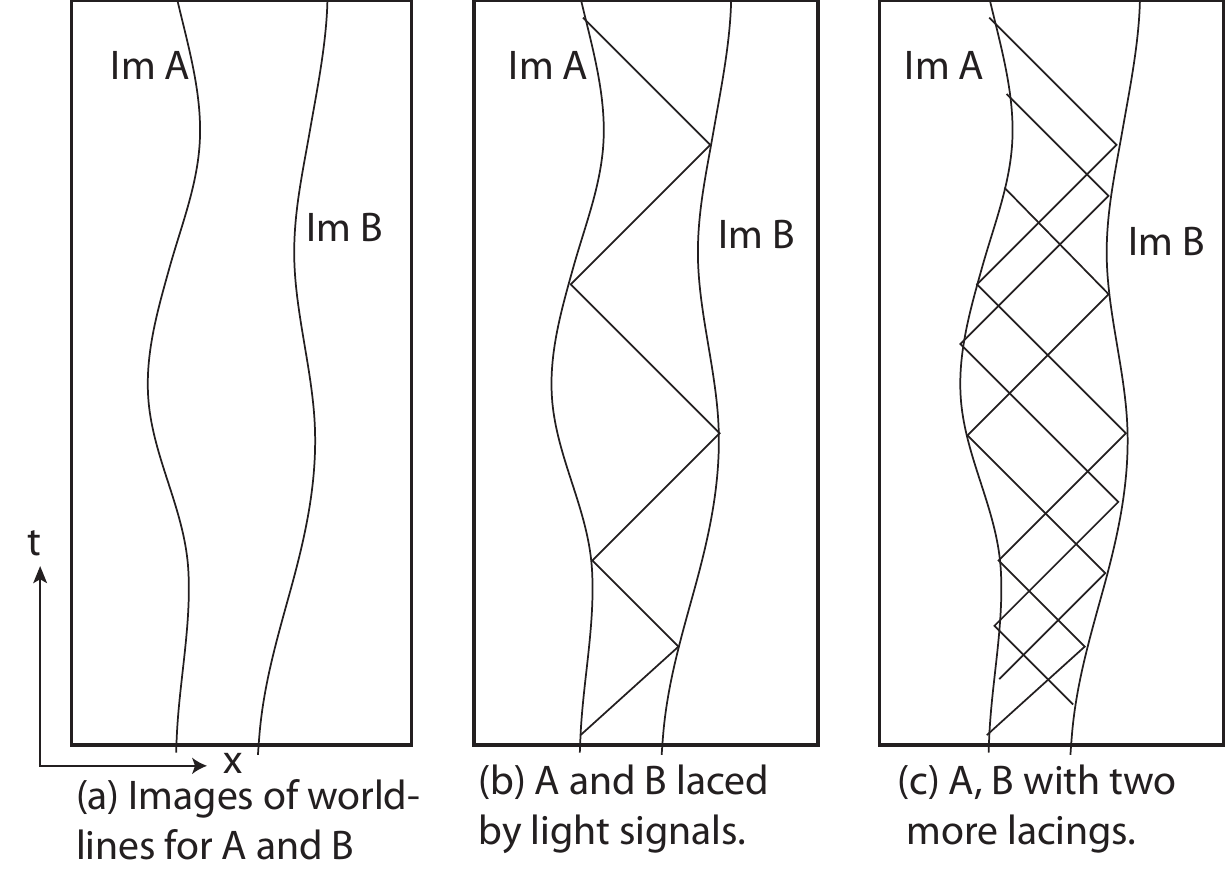}}
\caption{(a) Images of worldlines for open machines $A$ and $B$ freely
  chosen; (b) A lacing of light signals that defines tick
  events; (c) Interpolated lacings of light signals added to make
  $\Delta_{ABA}=\Delta_{BAB}=3$.}\label{fig:4}
\end{figure}

A different situation for two machines arises in case only the image of
$A$'s worldline is specified while its clocking left to be determined.  In
this case the image of any $B$ radar linkable to $A$ can be freely chosen,
after which the clocking of both $A$ and $B$ is constrained, as illustrated
in Fig.~\ref{fig:4} for the toy case of flat spacetime with 1 space
dimension.  To illustrate the constraint on clocking, we define a
``lacing'' of light signals to be a pattern of light signals echoing back
and forth between two open machines as illustrated in Fig.~\ref{fig:4}~(b).
For any event $a_0$ chosen in the image of $A$, there is a lacing that touches
it.  In addition to the choice of $a_0$, one can choose any positive
integer $N$ to be $\Delta_{ABA}$, and choose $N-1$ events in the image of
$A$ located after $a_0$ and before the next $A$-event touched by
the lacing of light signals.  The addition of lacings that touch each of
the $N-1$ intermediating events corresponds to a repeating channel $\abr$
with echo count $\Delta_{ABA}=N$, along with a repeating channel $\bar$
with the same echo count $\Delta_{BAB}=N$.  This construction does not
depend on the dimension of the spacetime nor on its flatness, and so works
also for a curved spacetime having the property of order preservation.

\subsection{Example of free choice characterized by a transformation group.}\label{sec:group}
Evidence of channels as patterns of clock readings leaves open a choice of
worldlines for its explanation.  In the preceding example of laced channels
between open machines $A$ and $B$, part of this openness can be reflected
within analysis by the invariance of the channels under a subgroup of the
group of clock adjustments that ``slides the lacings,'' as follows.
Suppose that transmissions of an open machine $A$ occur at given values of
$A$-readings.  We ask about clock adjustments that can change the events of
a worldline that correspond to a given $A$-reading.  If a clock adjustment
$f_A$ takes original $A$-readings $\zeta_A$ to a revised $A$-readings
$f_A(\zeta_A)$, transmission events triggered by the original clock
readings become triggered when the re-adjusted clock exhibits the
\emph{same readings}.  As registered by original readings, the adjusted
transmission occurs at $\zeta'_A=f_A^{-1}(\zeta_A)$.  Based on this
relation we inquire into the action of subgroups of $H\times H$ on the
readings of the clocks of two open machines $A$ and $B$.  In particular,
there is a subgroup $K(A,B) \subset H\times H$ that expresses possible
revisions of explanations that leave invariant the repeating channels with
constant echo count $N$.  An element $f_A\times f_B \in K(A,B)$ is a pair
of clock adjustments that leaves the channels invariant, and such a pair
can be chosen within a certain freedom.  For the adjustment $f_A$ one is
free to: (a) assign an arbitrary value to $f_A^{-1}(0)$; and (b), if $N>1$,
then for $j,k=1,\ldots,N-1$, choose the value of $f_A^{-1}(j)$ at will,
subject to the constraints that $k>j\Rightarrow f^{-1}(k)>f^{-1}(j)$ and
$f^{-1}(N-1)$ is less than the original clock reading for the re-adjusted
first echo from $f^{-1}(0)$.  With these choices, $f_B$ is then constrained
so that each lacing maps to another lacing.  The condition (a) slides a
lacing along the pair of machines; the condition (b) nudges additional
lacings that show up in the interval between a transmission and the receipt
of its echo.  In this way a freedom to guess within a constraint imposed by
evidence becomes expressed by $K(A,B)$.

\subsection{Channels among more than two open machines.}
Moving to more than two machines, we invoke the
\begin{quote}
  \textbf{Definition:} an \emph{arrangement of open machines} consists of
  open machines with the specification of some or all of the channels from
  one to another, augmented by proper periods of the clock of at least one
  of the machines.
\end{quote}
(Without specifying some proper periods, the scale of separations of
one machine from another is open, allowing the arrangement to shrink
without limit, thus obscuring the effect of spacetime curvature.)

Although gentle spacetime curvature has no effect on the possible
channels linking two open machines, spacetime curvature does affect
the possible channels and their echo counts in some arrangements of
five or more machines, so that the channels that can be implemented
are a measure spacetime curvature.
\begin{figure}[h]
\centerline{\includegraphics[width=4.9in]{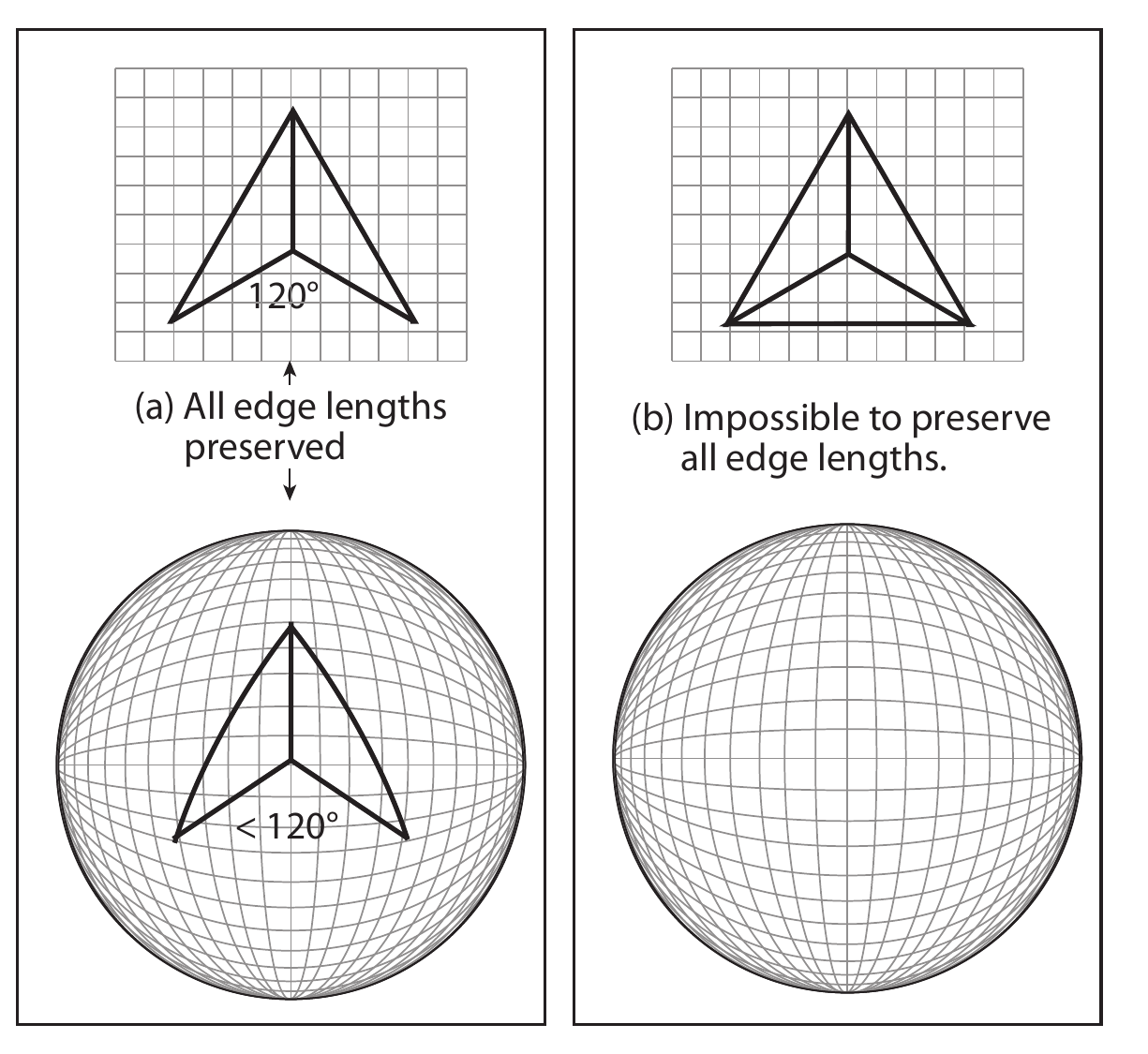}}
\caption{Plane figures, one of which maps to a sphere while preserving edge lengths}
\label{fig:sphere}
\end{figure}
The way that spacetime curvature affects the possible arrangements of
channels is analogous to the way surface curvature in Euclidean geometry
affects the ratios of the lengths of the edges of embedded graphs.  The
effect on ratios of edge lengths shows up in mappings from embeddings of
graphs in a plane to their images on a sphere.  For example, a triangle can
be mapped from a plane to a generic sphere, in such a way that each edge of
the triangle is mapped to an arc of the same length along a great circle on
the sphere.  The same holds for two triangles that share an edge, as
illustrated in Fig.~\ref{fig:sphere}, panel (a); however, the Gauss
curvature of the sphere implies that the complete graph on 4 vertices
generically embedded in the plane, shown in panel (b), cannot be mapped so
as to preserve all edge lengths.  The property that blocks the preservation
of edge ratios is the presence of an edge in the plane figure that cannot
be slightly changed without changing the length of at least one other edge;
we speak of such an edge as ``frozen.''

In a static spacetime, which is all we have so far investigated, a
generic arrangement of 4 open machines is analogous to the triangle
on the plane in that it can be mapped to any gently
curved spacetime in such a way as to preserve all the echo counts.
\begin{prop}\label{prop:nine}
Assume four open machines in a static spacetime, with one machine
stepped with a proper-time period $p_\tau$, and let $N$ be any
positive integer.  Then, independent of any gentle Riemann curvature
of the spacetime, the four open machines can be arranged, like
vertices of a regular tetrahedron, to have six two-way channels with
null phases, with all echo counts being $2N$.
\end{prop}
\textit{Proof:} Assuming a static spacetime, choose a coordinate system
with all the metric tensor components independent of the time coordinate,
in such a way that it makes sense to speak of a time coordinate distinct
from space coordinates (for example, in a suitable region of a
Schwarzschild geometry).  Let$V_1$ denote the machine with specified proper
period $p_{\tau}$, and let $V_2$, $V_3$, and $V_4$ denote the other three
machines.  For $i,j \in \{1,2,3,4\}$, $i \ne j$, we prove the possibility,
independent of curvature of the channels
\begin{equation}\label{eq:Vs}
\stackrel{\xrightarrow{\hspace*{0.8cm}}}{V_iV_j}=\{(k,k+N.0)|k \text{
  any integer}\}.  
\end{equation}
Let each of four machines be located at some
fixed spatial coordinate.  Because the spacetime is static, the coordinate
time difference between a transmission at $V_1$ and a reception at any
other vertex $V_j$ (a) is independent of the value of the time coordinate
at transmission and (b) is the same as the coordinate time difference
between a transmission at $V_j$ and a reception at $V_1$.  For this reason
any one-way repeating channel of the form (\ref{eq:Vs}) can be turned
around to make a channel in the opposite direction, so that establishing a
channel in one direction suffices.  For transmissions from any vertex to
any other vertex, the coordinate-time difference between events of
transmission equals the coordinate-time difference between receptions. A
signal from a transmission event on $V_1$ propagates on an expanding light
cone, while an echo propagates on a light cone contracting toward an event
of reception on $V_1$.  Under the constraint that the echo count is $2N$,
(so the proper duration from the transmission event to the reception event
for the echo is $2N p_{\tau}$), the echo event must be on a 2-dimensional
submanifold---a sphere, defined by constant radar distance $N p_{\tau}$
of its points from $V_1$ with transmission at a particular (but arbitrary)
tick of $V_1$.  In coordinates adapted to a static spacetime, this
sphere may appear as a ``potatoid'' in the space coordinates, with
different points on the potatoid possibly varying in their time coordinate.
The potatoid shape corresponding to an echo count of $2N$ remains constant
under evolution of the time coordinate.  Channels from $V_1$ to the other
three vertices involve putting the three vertices on the potatoid.
Put $V_2$ anywhere on the potatoid.  Put $V_3$ anywhere on the ring
that is intersection of potatoid of echo count $2N$ radiated from $V_2$ and
that radiated from $V_1$.  Put $V_4$ on an intersection of the potatoids
radiating from the other three vertices.\\ Q.E.D.

According to Prop \ref{prop:nine} the channels, and in particular the
echo counts possible for a complete graph of four open machines in
flat spacetime are also possible for a spacetime of gentle static
curvature, provided that three of the machines are allowed to set
their periods not to a fixed proper duration but in such a way that
all four machines have periods that are identical in coordinate time.
The same holds if fewer channels among the four machines are
specified.

But for five machines, the number of channels connecting them matters.
Five open machines fixed to space coordinates in a static spacetime are
analogous to the 4 vertices of a plane figure, in that an arrangement
corresponding to an incomplete graph on five vertices can have echo counts
independent of curvature, while a generic arrangement corresponding to a
complete graph must have curvature-dependent relations among its echo
counts.

\begin{prop}\label{prop:9.5}
Assuming a static spacetime, consider an arrangement of five open
machines obtained by starting with a tetrahedral arrangement of four
open machines with all echo counts of $2N$ as in
Prop. \ref{prop:nine}, and then adding a fifth machine: independent of
curvature, a fifth open machine can be located with two-way channels
having echo counts of $2N$ linking it to any three of the four
machines of tetrahedral arrangement, resulting in nine two-way
channels altogether.
\end{prop}
\textit{Proof:} The fifth machine can be located as was the machine
$V_4$, but on the side opposite to the cluster $V_1$, $V_2$, $V_3$.\\
Q.E.D.

\begin{figure}[h]
\centerline{\includegraphics[width=4.9 in]{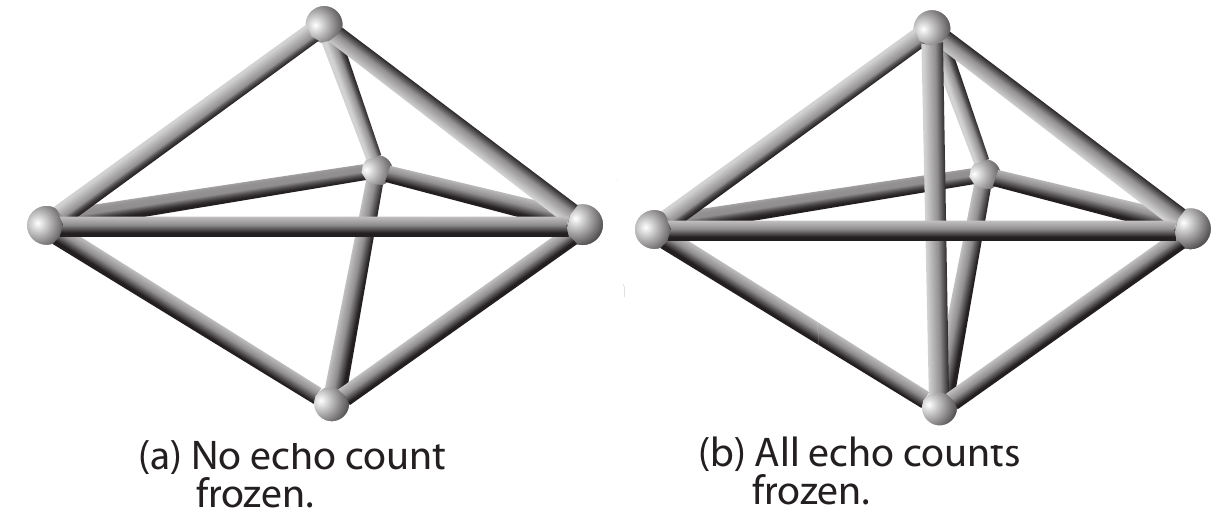}}
\caption{(a) 5 open machines with 9 two-way channels; (b) 
Five open machines with all 10 two-way channels}\label{fig:5pt}
\end{figure}

In contrast to an arrangement of 5 open machines having 9 two-way channels,
illustrated in Fig.~\ref{fig:5pt} (a) consider an arrangement analogous to
a complete graph on five vertices, having ten two-way channels, as
illustrated in Fig.~\ref{fig:5pt} (b).  For five open machines in a generic
spacetime, not all of the ten two-way channels can have the same echo
counts.  Instead, channels in a flat spacetime as specified below can exist
with about the simplest possible ratios of echo counts.  Label five open
machines, $A_1$, $A_2$, $A_3$, $B_1$, and $B_2$.  Take $B_1$ to be stepped
by a clock ticking at a fixed proper period $p_\tau$, letting the other
machines tick at variable rates to be determined.  Let $X$ be any machine
other than $B_1$.  For a flat spacetime it is consistent for the proper
periods of all 5 machines to be $p_\tau$, for the echo counts
$\Delta_{B_1XB_1}$ to be $4N$ and for the echo counts $\Delta_{A_i A_jA_i}$
to be $6N$, leading to the following twenty channels, conveniently viewed
as in Fig.~\ref{fig:5pt} (b) as consisting of ten two-way channels.
\begin{eqnarray}
\stackrel{\xrightarrow{\hspace*{0.7cm}}}{A_nB_j}&=&\{(k.0,k+2N.0)|k=0,1,2\ldots\}
\label{eq:ch51}\\
\stackrel{\xrightarrow{\hspace*{0.7cm}}}{B_jA_n}&=&\{(k.0,k+2N.0)|k=0,1,2\ldots\}
\label{eq:ch52}\\
\stackrel{\xrightarrow{\hspace*{0.7cm}}}{B_1B_2}&=&\{(k.0,k+2N.0)|k=0,1,2\ldots\}
\label{eq:ch53}\\
\stackrel{\xrightarrow{\hspace*{0.6cm}}}{B_2B_1}&=&\{(k.0,k+2N.0)|k=0,1,2\ldots\},
\label{eq:ch54}\\ \nonumber
&&\\
\stackrel{\xrightarrow{\hspace*{0.9cm}}}{A_nA_{n+1}}&=&\{(k.0,k+3N.0)|k=0,1,2\ldots\}\label{eq:ch55}\\
\stackrel{\xrightarrow{\hspace*{0.9cm}}}{A_{n+1}A_n}&=&\{(k.0,k+3N)|k=0,1,2\ldots\}\label{eq:ch56}
\end{eqnarray}
\begin{prop}\label{prop:ten}
Consider 5 open machines each fixed to space coordinates in a static curved
spacetime in which the machines are all pairwise radar linkable, with
10 two-way channels connecting each machine to all the others; then:
\begin{enumerate}
\item Allowing for the periods of the machines other than $B_1$ to
  vary, it is consistent with the curvature for all but one of the ten
  two-way channels (\ref{eq:ch51}--\ref{eq:ch56}) to have null phases
  and echo counts as in a flat spacetime, but at least one two-way
  channel must have a different echo count that depends on the
  spacetime curvature.
\item Suppose $m$ of the 10 two-way links are allowed to have non-zero
  phases.  If the spacetime does not admit all phases to be null, in
  generic cases the least possible maximum amplitude of a phase
  decreases as $m$ increases from 1 up to 10.
\item The periods of the clocks of the open machines can be taken to
  be the coordinate-time interval corresponding to the proper period
  $p_\tau$ of $B_1$.
\end{enumerate}
\end{prop}
\textit{Proof:} Reasoning as in the proof of Prop.\  \ref{prop:nine} with
its reference to a static spacetime shows that the same echo counts
are possible as for flat spacetime \emph{with the exception} that at
least one of the two-way channels must be free to have a different
echo count.  For $m < 10$, similar reasoning shows that allowing $m+1$
machines to vary in echo count allows reduction in the maximum variation
from the echo counts in a flat spacetime, compared to the case in
which only $m$ machines are allowed to vary in echo count.\\Q.E.D.

Adding the tenth two-way channel to an arrangement of five open
machines ``freezes'' all the echo counts.  To define \emph{freezing}
as applied to echo counts, first note an asymmetry in the dependence
of echo counts on clock rates.  Consider any two machines $A$ and
$B$. While $B$ can change the echo count $\Delta_{BAB}$ by changing
its clock rate, the echo count $\Delta_{ABA}$ is insensitive to $B$'s
clock rate.  An echo count $\Delta_{ABA}$ will be said to be \emph{to}
$B$ and \emph{from} $A$.
\begin{quote}
  \textbf{Definition:} An arrangement of open machines is
  \emph{frozen} if it has an echo count to a machine $B$ that cannot
  be changed slightly without changing another echo count to $B$.
\end{quote}
The property of being frozen is important because of the following.
\begin{prop}
  Whether or not a frozen arrangement of open machines is consistent
  with an hypothesized spacetime depends on the Weyl curvature of
  the spacetime.
\end{prop}

\subsection{Five open machines near Earth.}
Here is a quantitative example in which we think of the 5 open
machines linked as in a complete graph by ten two-way channels.  We
picture the 5 machines as carried by 5 space vehicles following
closely a radial geodesic in a Schwarzschild geometry corresponding to
the Earth as a central mass.  In this example the variation of echo
counts necessary to accommodate curvature is small enough to be
expressed by non-null phases of reception, without changing the
integer part of any echo count.  In Fermi normal coordinates centered
midway between the radially moving open machines $B_1$ and $B_2$ one
has the metric
\begin{eqnarray}\label{eq:5fnc}
  ds^2&=&-c^2[1+\mu(y^2+z^2-2x^2)]dt^2 \nonumber \\&-&\frac{2\mu}{3}
(xz\,dx\,dz+xy\,dx\,dy-2yz\,dy\,dz)\nonumber\\
&+&\left(1+\frac{\mu}{3}(y^2+z^2)\right)dx^2+
\left(1+\frac{\mu}{3}(x^2-2z^2)\right)dy^2\nonumber \\
&+&\left(1+\frac{\mu}{3}(x^2-2y^2)\right)dz^2,
\end{eqnarray}
where $\mu:=GM/(c^2r^3)$, $r$ is the Schwarzschild radial coordinate
to the origin of the Fermi normal coordinates, $x$ is the radial
distance coordinate from from the center point between $B_1$ and
$B_2$, and $y$ and $z$ are transverse to the radial geodesic
\cite{manasse}.  To work in SI units rather than the geometrized units of
\cite{manasse}, we write speed of light, $c$, explicitly.

We ignore the temporal variation of $r$ in comparison with the
dynamics of light signals between space vehicles, thus treating the
Fermi normal coordinates as pertaining to a static spacetime.  Locate
each of the 5 open machines at fixed values of $x,y,z$, as follows.
The metric (\ref{eq:5fnc}) is symmetric under rotation about the
$x$-axis.  Let $B_1$ and $B_2$ be located symmetrically at positive
and negative values, respectively, of the $x$-axis, and let $A_0$,
$A_1$, and $A_2$ be located on a circle in the plane $x=0$.  With the
five machines so located, the coordinate-time difference between
transmissions is then the same as the coordinate-time difference
between receptions, and the coordinate-time delay in one direction
equals that in the opposite direction (as stated in the proof of
Prop. \ref{prop:nine}).  We construct seven two-way channels as in
(\ref{eq:ch51}--\ref{eq:ch54}) with null phases and show that the
remaining 3 two-way channels can have the equal phases, but that this
phase $\phi$ must be non-null and dependent on curvature, as in
\begin{eqnarray}
\stackrel{\xrightarrow{\hspace*{0.9cm}}}{A_nA_{n+1}}
&=&\{(k.0,k+3N.\phi)|k=0,1,2\ldots\}\label{eq:ch55a}\\
\stackrel{\xrightarrow{\hspace*{0.9cm}}}{A_{n+1}A_n}
&=&\{(k.0,k+3N.\phi)|k=0,1,2\ldots\}\label{eq:ch56a}
\end{eqnarray}
\begin{prop}\label{prop:eleven}
Under the stated conditions, if the effect of curvature  is small enough so
that $27 GMN^3p_\tau^2/(4r^3) < 1$ then
\begin{equation}\label{eq:p11}
  \phi= -\frac{27GM N^3p_\tau^2}{8r^3}.
\end{equation}
\end{prop}
\textit{Proof}: (by calculation of $\phi$): 
\begin{enumerate}
\item Given $p_{\tau}$, determine coordinate-time period, denoted $p_t$, to
  first order in curvature.  This is the interval of coordinate time over a
  proper period $p_\tau$ of $B_1$ , which comes from evaluating $g_{00}$ at
  $x(B_1)$, with $x(B_1)$ evaluated at order zero:
\begin{equation}\label{eq:Dt}
p_t=(1+\mu N^2p_\tau^2c^2)p_\tau  
\end{equation}
\item Determine $x(B_1)$ to first order in curvature as the value such
that the coordinate time difference for a null geodesic from $B_2$ to
$B_1$ is $2Np_t$.  This leads to
\begin{equation}\label{eq:51}
  2Np_t=2\int_0^{x(B_1)}\frac{dt}{dx}dx,
\end{equation}
where $dt/dx$ is obtained from (\ref{eq:5fnc}) evaluated at
$y=z=dy=dz=0$.   Along the $x$-axis one finds $cdt/dx=1+\mu x^2$.
Substituting this into (\ref{eq:51}), integrating, and solving to
first order in curvature yields
\begin{equation}\label{eq:x1}
  x(B_1)= Np_t c\left(1-\frac{\mu}{3}N^2p_t^2c^2\right).
\end{equation}
\item Determine the radius of the circle on which $A_n$ lies,  $n = 0, 1, 2$.
The coordinates are symmetric under rotation about the $x$-axis,
so one can locate $A_0$ on the line $x=z=0$.  Then the radius
is just the coordinate $y(A_0)$.  The difference in coordinate time
between a null geodesic traversing from $B_1$ to $A_0$ and the
coordinate time for a null trajectory that is linear in the
coordinates is zero to first order in curvature.  Thus, to first order,
the coordinate time difference is that of a null curve following
$z=0$, $y=(x_1-x)y_0/x_1$, $dy= -(y_0/x_1)dx$, where we write $y_0$ for
$y(A_0)$ and $x_1$ for $x(B_1)$.  Eq.~(\ref{eq:5fnc}) implies for this
null curve:
\begin{equation}
c^2\!\left[1+\mu\left(\!(x_1-x)^2\frac{y_0^2}{x_1^2}-2x^2\!\right)\!\right]dt^2=
 \frac{x_1^2+y_0^2}{x_1^2}\left(1+
    \frac{\mu x_1^2y_0^2}{3(x_1^2+y_0^2)}\right)dx^2,
  \end{equation}
leading to the relation to first order in curvature:
\begin{equation}
 c \frac{dt}{dx}=
\frac{\sqrt{x_1^2+y_0^2}}{x_1}\left(\!1+\frac{\mu
  x_1^2y_0^2}{6(x_1^2+y_0^2)}\right)\!
\left[1-\frac{\mu}{2}\left((x_1-x)^2\frac{y_0^2}{x_1^2}-2x^2\right)\right].
\end{equation}
Integration gives
\begin{equation}\label{eq:t1}
  2Np_t c=\sqrt{x_1^2+y_0^2}\left[1+\frac{\mu}{6}
\left(\frac{x_1^2y_0^2}{x_1^2+y_0^2}+2x_1^2-y_0^2\right)\right].
\end{equation}
Substituting (\ref{eq:Dt}) and (\ref{eq:x1}) into (\ref{eq:t1}) and
solving to first order in $\mu$ for $y_0$ yields
\begin{equation}
  y_0\equiv y(A_0)=\sqrt{3}Np_t c\left(1+\frac{\mu}{8} N^2p_t^2c^2\right)
  +O[(\mu N^2p_t^2c^2)^2].
\end{equation}
\item Determine the coordinate-time delay for transmission from $A_0$
  to $A_1$, which is the same coordinate-time delay for all the
  transmissions between $A_m$ and $A_n$, $n\ne m$, for $n,m= 0,1,2$.
  The metric is symmetric under rotation about the $x$-axis, which
  allows us to rotate $A_0$ to the position $x=0$, $z=y_0/2$,
  $y=-(\sqrt{3}/2)y_0$, and to locate $A_1$ at $x=0$, $z=y_0/2$,
  $y=(\sqrt{3}/2)y_0$.  Again, deviations of null geodesic from the
  linear relation between coordinates make zero first-order
  contribution, so we compute the coordinate-time delay for a
  null-curve from $A_0$ to $A_1$ along the line $x=0$, $z=y_0/2$,
  which to first order in curvature is:
\begin{equation}
  t(A_1 \text{rec})-t(A_0 \text{x-mit})=3Np_t (1-9\mu N^2p_t^2c^2/8) 
\end{equation}
\item Converting from coordinate-time delay to the coordinate-independent
echo count, under the hypothesis that $27\mu N^3p_t^2c^2/8<1/2$, we
arrive at the fractional part of $t(A_1 \text{rec})-t(A_0
\text{x-mit})$ being $\phi$ as stated in the Prop.~\ref{prop:eleven}.
\end{enumerate}
Q.E.D.\vspace*{5pt}

\noindent Note that the proper periods of both $B_1$ and $B_2$ are
$p_\tau$, while, to first order in curvature, that of the $A_n$ is
$p_\tau(1-\mu N^2p_\tau^2c^2)$.

\subsection{Changing curvature limits bit rate.}
The dependence of echo counts on curvature has an interesting
implication.  When channels are to be maintained in the face of varying
curvature, or in cases where there is uncertainty about what curvature
describes their situation, the variability in curvature imposes a
lower bound on clock rates and hence an upper bound on the rate at
which information can be transmitted from one open machine to another.
For the situation of the preceding example, this limit is readily
obtained, as follows.  For simplicity, assume that the positions and
clock rates are continually adjusted to maintain null phases for all
but the three channels
$\stackrel{\xrightarrow{\hspace*{0.9cm}}}{A_nA_{n\pm 1}}$.  From
(\ref{eq:ch53}) we have that $L\approx 2Np_\tau c$, which with
Prop. \ref{prop:eleven} and the fact that $p_t\approx p_\tau$ implies
$\phi\approx - 27ML^3/(32r^3c^3p_{\tau})$, which with (\ref{eq:main})
implies
\begin{equation}\label{eq:pcurve}
  p_{\tau}>\frac{27 G M L^3}{32r^3c^3}
\end{equation}
Suppose the cluster of 5 open machines is arranged to have the
proper radar distance $L$ from $B_1$ to $B_2$ be 6,000 km, and suppose the
cluster descends from a great distance down to a radius of $r=30,000$ km
from an Earth-sized mass $M_{\oplus} = 5.98\times 10^{24}$ kg.  
With these values of the parameters, for the phases for
the channels $\stackrel{\xrightarrow{\hspace*{0.9cm}}}{A_nA_{n\pm 1}}$
to satisfy (\ref{eq:main}), it is necessary that
\begin{equation}
  p_\tau> 1.0\times 10^{-13} \text{ s}.
\end{equation}
In case an alphabet conveys $b$ bits/character, the maximum bit rate
for all the channels in the 5-machine cluster is
$b/p_\tau< 10^{13}b$ bits/s.

\section{Steering while listening to the
  unforeseeable}\label{sec:adj}

The preceding section displays ``blackboard models'' of clocks,
expressed in the mathematical language of differential geometry.
Turning from Type-I questions to questions of Type II, we now look at
how such models get put to work when they are encoded into programs of
computers that steer open machines in clock rate and in position
toward aiming points.  For questions of Types II and III, besides the
models that explain or predict evidence, the evidence itself comes
into play.  Models taking part in the steering of physical clocks
contribute to the generation of echo counts as evidence that, one
acquired, can stimulate the guessing of new models that come closer to
the aiming point.

\subsection{Importing quantum uncertainty into general relativity.}
To express the effect of quantum uncertainty on deviations from aiming
points, one has to introduce quantum uncertainty into the
representation of clocks by general-relativistic worldlines.  This
introduction hinges on the ever-crucial distinction between evidence
and its explanations.  Timelike worldlines and null geodesics in
explanations, being mathematical, can have no \emph{mathematical}
connection to physical atomic clocks and physical signals. To make any
(non-mathematical) connection, one has to invoke the logical freedom
to make a guess.  Within this freedom, without logical conflict, one
can interpret events of signal reception as corresponding to
expectation values in the sense of quantum theory.  This
intermediating layer of modeling explains some of the deviations of an
atomic clock from an imagined proper clock, represented as a
worldline.  This is no ``unification'' of quantum theory and the
theory of general relativity, merely a recognition that both theories
are blackboard systems of explanation, distinct from evidence, and
that pieces from one can, under certain circumstances, be joined to
the other.

\subsection{Need for prediction in steering toward an aiming point.}
For reasons that include quantum uncertainty, coming close to an
aiming point stated in terms of channels and a proper frequency scale
requires steering.  In steering, evidence of deviations from the
aiming point combine with hypotheses concerning how to steer
\cite{medterm,algorithm}.  For example, consider a case of an aiming
for two open machines $A$ and $B$, as in the first example of
Sec.~\ref{sec:typeI}.  Recall that the open machine $A$ is modeled by
a given worldline with given clock readings$\zeta_A$.  Machine $B$
aims to maintain a two-way, null-phase channel of given $\Delta_{ABA}=
\Delta_{BAB}$.  To this end $B$ registers arriving phases of reception
and adjusts its clock rate more or less continually to keep those
phases small.  But $B$ also needs to steer in position.  Deviations in
$B$'s position show up as phases of echoes registered by $A$, so the
steering of machine $B$ requires information about receptive phases
measured by $A$.  The knowledge of the deviation in position of $B$ at
$\zeta_B$ cannot arrive at $B$ until its effect has shown up at $A$
and been echoed back as a report to $B$, entailing a delay of at least
$\Delta_{BAB}$, hence requiring that machine $B$ predict the error to
which its steering responds.  Machine $B$ must predict ahead by at
least $\Delta_{BAB}$.  That is, steering deviations by one open
machine are measured in part by their effect on receptive phases of
other open machines, so that steering of one machine requires
information about receptive phases measured by other machines, and the
deviations from an aiming point must increase with increasing
propagation delays that demand predicting further ahead.

As is clear from the cluster of five machines discussed in
Sec.~\ref{sec:typeI}, the aiming-point phases cannot in general all be
taken to be zero.  For any particular aiming-point phase $\phi_0$
there will be a deviation of a measured phase quantity $\phi$ given by
\begin{equation}
  \delta:= \phi-\phi_0
\end{equation}
Whatever the value of $\phi_0$, adjustments to contain phases within
tolerable bounds depends on phase changes happening only gradually, so
that trends can be detected and responded to on the basis of adequate
prediction. \vspace{6pt}\\
\noindent\textbf{Remarks:}
\begin{enumerate}
\item While it is often convenient to assume that cycle counts of open
machines are free of uncertainty, recognizing uncertainty in measured
phases and their deviations from aiming point has an immediate and
interesting implication.  For logic to work in a network, transmission
of logical symbols must preserve sharp distinctions among them; yet
the maintenance of sharp distinctions among transmitted symbols
requires responses to fuzzy measurements.
 \item The acquisition of logical synchrony in digital communications
   involves an unforeseeable waiting time, like the time for a coin on
   edge to fall one way or the other \cite{meyr,1639}.
\end{enumerate}

\subsection{Adjusting the aiming point.}\label{sec:adjAim} 
Here we touch on questions of Type III.  Up to this point we have
looked at one or another manifold with metric $(M,g)$ as some given
hypothesis, whether explored on the blackboard or coded into an open
machine to serve in steering toward an aiming point.  For it use in
steering we think of $(M,g)$ as ``given''---one might say for use
``ballistically,'' without any ``piloting.''  But Type-III questions
recognize that unforeseen deviations of phases outside of tolerances
can happen, and an aiming point based on a hypothesized metric tensor
can be found to be unreachable.  Recognizing that an hypothesis of a
metric tensor can reach the end of its useful life calls for
``piloted'' hypothesis making, recognizing that each hypothesis of a
metric tensor field is provisional, to be revised as prompted by
deviations outside allowed tolerances assigned for steering toward an
aiming point that incorporates that metric tensor field.

Drawing on measured phases as evidence in order to adjust a hypothesis
of a metric tensor is one way to view the operation of the Laser
Interferometer Gravitational-Wave Observatory (LIGO) \cite{ligo}.
While LIGO sensitivity drops off severely below 45 Hz, the arrangement
of five open machines of Prop.~\ref{prop:ten} has no low-frequency
cutoff, and so has the potential to detect arbitrarily slow changes in
curvature.

\section{Discussion}\label{sec:discussion}
In the physics of Newton, ``time'' as a concept offers a future
that flows downstream to the present and into the past.  In special
relativity Einstein grounds a concept of time on `time local to a
clock', spread out by the synchronization of separated clocks
(assuming no drift).  But except locally, Einstein synchronization is
unavailable to the curved spacetimes of general relativity
\cite{perlick}.  Still, general relativity holds fast to the image of
a predictable future, seemingly unwelcoming of surprise, as if what I
do not foresee within my (relativistic) future is the fault of my
ignorance, which I can hope to remedy.  But if explanations in terms
of wave functions are undetermined by any amount of evidence, and
therefore are subject to surprises that prompt their revision, how do
we float the future on any `river of time'?

By accepting unforeseeable events, and by reflecting within itself a
range of guesses that respect given evidence, logical synchronization
as a discipline within physics gives up the hope of using prediction
to evade the need for adjustment, and instead employs predictions,
dependent on guesswork, to define the departures that call for
adjustments that respond to unforeseeable events, welcoming surprises
that lead to the revision of predictions.  By giving up predictability
in favor of adjustment as fundamental to physics, logical
synchronization opens avenues of inquiry invisible under any concept
of `time' that fails to acknowledge the foreseeable.  Here are three
examples:
\begin{enumerate}  
\item \textit{How freedom of choice of wave function for atomic clocks
implies freedom of choice in the smoothing of a metric tensor as an
explanation of clock readings.}  Both for arrangements of clocks near and on Earth
and for cosmological endeavors, one has occasion to choose a metric
tensor field consistent with one or another body of experimental
evidence of clock readings in the form of Fig.~\ref{fig:1}.  Clock readings come
with error bars on phases contribute evidence used in choosing a
metric tensor field.  Think of two error bars separated by a
variable distance; as the distance shrinks the range of slopes of
lines consistent with error bars increases. The resulting indeterminacy in
the metric tensor thus involves a scale over which smoothing is
applied, and at short scales of time and distance the room for choice
of both spacetime curvature and clock-rate variation grows while
maintaining consistency with experimental data.  \vspace*{5pt}
\item \textit{How chosen wave functions for atomic clocks affect
  spacetime curvature.}  Additional freedom of metric tensor fields
  chosen to explain clock readings arises because, as sketched in
  \ref{sec:clk}, a chosen wave function references the SI second to an
  imagined resonance at absolute zero temperature.  One might
  hypothesize a fine-scale variation in the 0 K offset of atomic
  clocks from from moment to moment and clock to clock as what we have
  called a clock-adjustment field \cite{aop84}, or one might smooth
  this adjustment field.  This difference between the smoothed and
  unsmoothed clock-adjustment fields implies a difference in the
  corresponding metric tensor fields chosen to explain clock readings.
  If the smoothed readings underlie an hypothesized metric tensor field
  $g$, then the unsmoothed readings would correspond to a metric
  tensor field as expressed by a conformal transformation
  $\tilde{g}=(1+\epsilon(x))g$, where $|\epsilon| \ll 1$ and $x$ is
  the 4-dimensional spacetime coordinate of some assumed chart.  The
  two metric tensor fields, even if differing only slightly, can imply
  markedly different spacetime curvatures because towo powers of
  derivatives---e.g. a term $(\nabla_a
  \epsilon)(\nabla^a\epsilon)$---that appears in relating the
  curvature of one spacetime to that of the other \cite{wald446}.
  Thus the smallness of the clock shift $\epsilon$ is multiplied in
  its effect on curvature by the rapidity of its variation as
  expressed in these derivatives.

\item \textit{Further development of group representations of freedom
  of choice in explanations of given evidence.}  As we have seen
  above, unforeseeability brings open choices of explanations that
  respect given evidence, such as the choice of a wave function,
  resolved by guesswork.  Guesses, by definition, cannot be decided by
  analysis, but in Sec. \ref{sec:group} one choice open to guesswork
  was characterized by a transformation group.  What else is there to
  do along these lines?
\end{enumerate}

\appendix
\section{Guesswork in atomic clocks stemming from the gap between
evidence and explanation as reflected in quantum theory}
\label{sec:A}

The operation of an atomic clock depends on a wave function chosen to
explain evidence attributed to the atom or atoms of the clock as these
are radiated by the oscillator.  How, though, is this wave function
chosen?  One learns quantum mechanics starting `the other way': given a
wave function and a positive operator-valued measure (POVM) expressing
a measurement procedure, one learns to calculate probabilities of
outcomes.  But an atomic clock poses an `inverse problem' of arriving
at a quantum model from given probabilities.  If one knew the Hilbert
space and a sufficient set of measurement operators, the density
operator (or the wave function) could be determined, provided that the
quantum state under study could be prepared repeatedly
\cite{fano,christandl}.  Assuming knowledge of measurement operators,
etc., this determination, called ``quantum tomography,'' is effective
in some cases \cite{feedbackBackAct}; however, no experiment can
determine a Hilbert space, and to determine the measurement operators
one has to determine the effects of the measuring procedures that they
express on laboratory preparations for which the density operators on
the assumed Hilbert space are known.  Thus the logic of quantum
tomography is circular, which raises the question: given probabilities
abstracted from laboratory work, how is one to arrive at a wave
function without assuming knowledge of measurement operators that
itself presupposes knowledge of some density operators?

Quantum theory reflects within the grammar of its mathematical
language a distinction between experimental results and their
explanations.  This distinction makes it possible to picture an
experimental set-up without presupposing any of its possible
explanations in terms of quantum states and measurement operators.
Picture an experimenter transmitting commands to an experimental
set-up and receiving reports of detections via a process-control
computer expressed by an open machine \cite{ams02}.  The commands act
set values of device parameters which we think of as knobs controlling
devices.  We call a list of knobs with their possible settings a
\emph{knob domain}, and a family of related experiments, some
involving more or different knobs than others corresponds to a lattice
of knob domains ; similarly one defines a lattice of detector domains
\cite{tyler07}.  Given a knob domain $\bm{K}$ and a detector domain
$\bm{\Omega}$, evidence to be explained quantum mechanically consists
of a \emph{parametrized probability measure} (PPM) which is a function
$\alpha\colon\bm{K}\times\bm{\Omega}\rightarrow [0,1]$, with the
property that \begin{equation} (\forall k\in
  \bm{K})\;\alpha(k,-)\colon\bm{\Omega}\rightarrow \,[0,1] \text{ is a
    probability measure on }\bm{\Omega}, \end{equation} so that the
PPM can be viewed as a function from a knob domain to probability
measures on $\bm{\Omega}$.  (Any of several metrics defined on spaces
of probability distributions induce a topology on a knob domain
\cite{tyler07}.  Lifts from probability distributions to PPMs allow
the definition of ``metric deviation'' as a quantitative difference
between two PPMs that have the same knob domain but can differ in
their detector domains \cite{tyler07}.)

Turning from evidence to its explanation, quantum language is used in
various formulations to explain or to predict a PPM with domains
$\bm{K}$ and $\Omega$.  For example, assume that the knob domain
splits into two pieces, $\bm{K}=\bm{K}_{\rm prep}\times \bm{K}_{\rm
  meas}$, the first for the knobs by which a density operator is
selected, the second for the knobs by which a Positive Operator-Valued
Measure (POVM) is selected.  A quantum-theoretic \emph{model} of the
PPM can be formulated as a structure
$(\bm{K},\bm{\Omega},\mathcal{H},\rho,M)$ with:
\begin{enumerate}
\item a Hilbert space $\mathcal{H}$,
\item  a function $\rho\colon\bm{K}_{\rm prep}\rightarrow 
\{\text{density operators on} \mathcal{H}\}$,
\item a function $M\colon\bm{K}_{meas}\rightarrow \{$POVMs on
  $\bm{\Omega}$ assigning positive operators on $\mathcal{H}$\}
\end{enumerate}
(Other formulations include unitary time evolution explicitly
\cite{aop05}.)
Models of a PPM $\alpha$ express $\alpha$ through the trace as a functor:
\begin{equation}
  \alpha(k,\omega)=\text{tr}[\rho(k)M(k,\omega)]
\end{equation}
Metric deviations are defined not only for PPMs but also for density
operators belonging to different models which can differ in their
Hilbert spaces and for POVMs that can differ in their detector
domains. When two models exhibit a positive metric deviation, we say
the models are metrically distinct. (In contrast, unitarily equivalent
models are \emph{not} metrically distinct.)

With respect to the `inverse problem' of choosing a model to explain
given probabilities, quantum theory enables the proof of the following:
 \begin{quote} 
\textbf{Proposition A.1}: For any given PPM there is an infinite set
of models, all metrically distinct \cite{aop05,tyler07,CUP}.
\end{quote} 
\begin{quote} 
\textbf{Proposition A.2}: Any two metrically distinct models of a PPM
can be extended in their detector domains or their knob domains to
imply metrically distinct extended PPMs \cite{aop05,tyler07}.
\end{quote} 
Thus many quantum models, involving distinct wave functions and POVMs,
map via the trace to any given PPM, but conflict with one another in
their implied probabilities for experiments that extend the given PPM.
so that choosing a wave function or a linear operator to explain an
experiment requires reaching beyond the confines of logic based on
evidence, and there is provably always the possibility that newly
acquired evidence will call for a change of mind.
With the recognition of unforeseeable events implicit in the use of
wave functions encoded into clocks, we see quantum theory as a
language in which one can think, speak, and trace the logic of
conclusions to assumptions, revising the assumptions when that is
called for.\vspace*{6pt}

\noindent\textbf{Remark:}
The narrow view of quantum language as expressing evidence by
probabilities precludes the application of quantum theory to feedback
that reacts promptly to individual occurrences of outcomes. Following
common practice, we adopt a wider view by admitting occurrences of
outcomes, not just their probabilities, into the language of quantum
theory, enabling the discussion of feedback that responds to outcomes
of individual occurrences of measurements.

Quantum cryptography offers two related examples of ambiguity in the
choice of wave functions, in which the issue is the security of
quantum key distribution against undetected eavesdropping.  In both
examples the wave function that naive tomography produces seriously
misleads; in effect Occam's razor fails.  An analysis of security that
assumes a low-dimensional Hilbert space overlooks vulnerabilities from
physical effects exploited by expanding the detector domain in such a
way that a larger Hilbert space along with a larger detector domain is
required to express these effects, as has been reported in an example
\cite{acin}; a more physical example involves the propagation of
pulses of polarized light generated by a set of four lasers.  A
popular model that expresses a density operator only with respect to
light polarization asserts security, but an enveloping model in which
the density operator also expresses variations in wave lengths emitted
by the lasers displays security vulnerabilities \cite{CUP}.


\begin{thebibliography}{99}
\bibitem{ams02} J. M. Myers and F. H. Madjid, ``A proof that measured data and
  equations of quantum mechanics can be linked only by guesswork,'' in
  S. J. Lomonaco Jr. and H.E. Brandt (Eds.) \textit{Quantum Computation and
    Information}, Contemporary Mathematics Series, vol. 305, American
  Mathematical Society, Providence, 2002, pp. 221--244.
\bibitem{aop05} F.~H. Madjid and J.~M. Myers, 
Ann.\ Physics \textbf{319} (2005), 251.
\bibitem{tyler07} J.~M. Myers and F.~H. Madjid, ``Ambiguity in
  quantum-theoretical descriptions of experiments,'' in
  K. Mahdavi and D. Koslover, eds., \textit{Advances in Quantum
    Computation}, Contemporary Mathematics Series, vol.~482
  (American Mathematical Society, Providence, I, 2009),
  pp.\ 107--123.
\bibitem{CUP} J. M. Myers and F. H. Madjid, ``What probabilities
  tell about quantum systems, with application to entropy and
  entanglement,'' in A. Bokulich and G. Jaeger, eds., \textit{Quantum
    Information and Entanglement}, Cambridge University Press,
  Cambridge UK, pp. 127--150 (2010).
\bibitem{hinkley} N. Hinkley, J. A. Sherman, N. B. Phillips,
  M. Schioppo, N. D. Lemke, K. Beloy, M. Pizzocaro, C. W. Oates, and
  A. D. Ludlow, 
Science \textbf{341} (2013) 1215.
\bibitem{allan87} D. W. Allan,
IEEE Transactions on Ultrasonics, Ferroelectrics, and
  Frequency Control, \textbf{UffC-34} (1987), 647.
\bibitem{opClk2005} P. Gill,
  Metrologia \textbf{42} (2005), S125.
\bibitem{opClk}C. W. Chou, D. B. Hume, T. Rosenband, and
  D. J. Wineland,  Science
  \textbf{329} (2010), 1630.
\bibitem{F1_2002} S. R. Jefferts et al., Metrologia \textbf{39} (2002), 321.
\bibitem{F1_2005} T. P. Heavner et al.,  Metrologia \textbf{42} (2005), 411.
\bibitem{F1op2005}T. E. Parker, S. R. Jefferts, T. P. Heavner, and E. A. Donley,
 Metrologia, \textbf{42} (2005), 423.
\bibitem{F2} T. E. Heavner et al. ``NIST F1 and F2'', 
Proc. 42nd Annual Precise Time and Time Interval Systems and
Applications Meeting, Reston VA (2010). 457.  Available at 
http://tf.boulder.nist.gov/general/pdf/2500.pdf.
\bibitem{jentschura} U. D. Jentschura, M. Puchalski, and
  P. J. Mohr, Phys. Rev. A \textbf{84} (2011), 064102.
\bibitem{DopFountain} J. Gu\'ena, R. Li, K. Gibble, S. Bize, and
  A. Clairon,   Phys. Rev. Lett. \textbf{106} (2011), 130801.
\bibitem{thermal} K. Beloy, U. I. Safronova, and A. Derevianko,
  Phys. Rev. Lett. \textbf{97} (2006), 040801 .
\bibitem{sp330} B. N. Taylor and A. Thompson, Eds, \textit{The
  International System of Units (SI)}, NIST Special Publication
  330 -- 2008 Edition, National Institutes of Science and
  Technology. 
\bibitem{turing} A.~M. Turing, 
Proc.\ London Math.\ Soc., Series 2, \textbf{42} (1936), 230.
\bibitem{soffel03} M. Soffel et al.,  The
  Astronomical Journal \textbf{126} (2003), 2687.
\bibitem{holtOcc} A.~W. Holt, ``Introduction to occurrence systems,'' in
  E.~L. Jacks, ed., \textit{Associative Information Techniques}, Proceedings of
  the Symposium held at General Motors Research Laboratories, 1968 (American
  Elsevier, New York, 1971), pp.~175--203.
\bibitem{mk_graph}F. Commoner, A.~W. Holt, S. Even, and A. Pnueli, ``Marked Directed Graphs,''
Journal of Computer and System Sciences \textbf{5}(5), 511--523 (1971).
\bibitem{peterson81} J.~L. Peterson, \textit{Petri Net Theory and the
    Modeling of Systems} (Prentice-Hall, Englewood Cliffs, NJ, 1981).
\bibitem{meyr} H. Meyr and G. Ascheid, \textit{Synchronization in
  Digital Communications}, Wiley, New York, 1990.
\bibitem{perlick} V. Perlick, ``On the radar method in
  general-relativistic spacetimes,'' in H. Dittus, C. L\"ammerzahl,
  and S. Turyshev, eds., \textit{Lasers, Clocks and Drag-Free
    Control: Expolation of Relativistic Gravity in Space}, (Springer,
  Berlin, 2008); also arXiv:0708.0170v1.
\bibitem{ligo} The LIGO Scientific Collaboration (http://www.ligo.org)
Rep. Prog. Phys. \textbf{72}, 076901 (2009)
\bibitem{manasse} F. K. Manasse and C. W. Misner, J. Math Phys., \textbf{4}
(1963) 735.
\bibitem{medterm} T. E. Parker, S. R. Jefferts, and T. P. Heavner,
  ``Medium-term frequency stability of hydrogen masers as measured by
  a cesium fountain,'' 2010 IEEE International Frequency Control
  Symposium (FCS), (2010) 318--323. (available at
  http://tf.boulder.nist.gov/general/pdf/2467.pdf)
\bibitem{algorithm} J. Levine and T. Parker, ``The algorithm used to
  realize UTC(NIST),'' 2002 IEEE International Frequency Control
  Symposium and PDA Exhibition (2002) 537--542.
\bibitem{1639} J. M. Myers and F. H. Madjid, ``Rhythms of Memory and
  Bits on Edge: Symbol Recognition as a Physical Phenomenon,''
  arXiv/1106.1639, 2011.
\bibitem{aop84} H. Madjid and J. M. Myers,  Ann.\ Physics \textbf{158}
  (1984), 421.
\bibitem{wald446} R. M. Wald, \textit{General Relativity}, University
  of Chicago Press, Chicago, 1984, p. 446.
\bibitem{fano} U. Fano, 
  Rev. Mod. Phys. \textbf{29} (1957), 74.
\bibitem{christandl} M. Christandl and R. Renner,
  Phys. Rev. Lett. \textbf{109} (2012), 120403.
\bibitem{feedbackBackAct} M. Hatridge et al.,  Science \textbf{339} (2013),
  178. 
\bibitem{acin} A. Ac\'in, N. Gisin, and L. Masanes, 
  Phys. Rev. Lett. \textbf{97} (2006), 120405.
\end{thebibliography}
\end{document}